\documentclass[aps,superscriptaddress,twocolumn,showpacs]{revtex4}

\usepackage[dvipdfm]{graphicx}
\usepackage{amsmath, amssymb}

\usepackage{color}

\usepackage{ulem}

\begin{document}
\title{Advection of passive particles over flow networks} 
\author{Shigefumi Hata}
\affiliation{Department of Physical Chemistry, Fritz Haber Institute of the Max Planck Society, Faradayweg 4-6, 14195 Berlin, Germany.}
\author{Hiroya Nakao}
\affiliation{Department of Mechanical and Environmental Informatics, Tokyo Institute of Technology, Ookayama 2-12-1, 152-8552 Tokyo, Japan.}
\author{Alexander S. Mikhailov}
\affiliation{Department of Physical Chemistry, Fritz Haber Institute of the Max Planck Society, Faradayweg 4-6, 14195 Berlin, Germany.}


\begin{abstract}
The problem of stochastic advection of passive particles by circulating conserved flows on networks is formulated and investigated.
The particles undergo transitions between the nodes with the transition rates determined by the flows passing through the links.
Such stochastic advection processes lead to mixing of particles in the network and, in the final equilibrium state, concentration of particles in all nodes become equal.
As we find, equilibration begins in the subset of nodes, representing flow hubs, and extends to the periphery nodes with weak flows.
This behavior is related to the effect of localization of the eigenvectors of the advection matrix for considered networks.
Applications of the results to problems involving spreading of infections or pollutants by traffic networks are discussed.
\end{abstract}



\maketitle

Advection phenomena play an important role in physics, biology, engineering and earth sciences~\cite{Pasquill1983, Mafraneto1994, Atena1995, Shraiman2000}.
They yield a general mechanism by which particles (pollutants) become spread over the system by conserved fluid flows.
Generally, the flow pattern is given and not influenced by variations in particles' concentration.
Statistical aspects of advection of a passive scalar by turbulent flows have attracted much attention~\cite{Shraiman2000}.

Advection processes have been extensively studied for continuous media, but there are also many situations where flows are passing over connections between descrete nodes constituting a network.
Obvious examples are provided by pipeline networks, used for delivery of gas or oil to a set of destinations, but the models of flow networks have been used also in systems biology when signal transduction effects were considered (see, e.g.~\cite{Kaluza2012}).
In the transportation context, traffic flows are established by trains, ships or aircraft on a regular scheduled service between rail stations, ocean harbours or airports.

Fluid flows in pipelines are conserved, so that the total amount of fluid, arriving to a redistribution node, is equal to the amount of fluid which leaves it.
Often, this is also true for the trafic flows where the number of carriers (ships or airplanes) entering a transportation node (a harbour or an airport) is, on time average, the same as the number of carriers departing from it.

Pipelines generally have source nodes, where the fluid is pumped into a network, and sinks where the fluid is taken away from it.
In contrast to this, traffic networks would typically have no nodes where the new transportation carriers are persistently created or existing carriers are persistently removed.
This means that the sources and sinks are then absent and steady patterns of circulating flows are maintained.

Stochastic transport of particles, such as pollutants or infectious agents, over networks can be described in terms of random Markov processes~\cite{Barrat2008}.
Diffusion processes on networks are of fundamental importance for spreading of infectious diseases~\cite{Satorras2001, Hufnagel2004, Colizza2006, Colizza2007} and dispersal connections between ecological habitats may significantly affect the dynamics and stability of a metapopulation~\cite{Hanski1998, Hata2013} (see also~\cite{Nakao2010, Kouvaris2012, Wolfrum2012}).
When modeling such phenomena, it is usually assumed that probabilities of transitions between the nodes are not correlated and, in principle, they can be arbitrarily assigned.  

In the present study, we consider the problem of advection, i.e.~of stochastic transport of particles by conserved circulating flows on networks.
We assume that the flow pattern is stationary and, for each node, total incoming and outgoing flows are equal.
The particles can be only transported together with a flow, so that the probability of transition from one node to another is proportional to the intensity of the flow passing through the respective link. 
As we show, flow conservation has strong implications for transport behavior.
At equilibrium, concentrations of particles in all nodes (with non-vanishing passing flows) are the same and, thus, the steady state is always uniform.
Equilibration of particle concentrations begins in the subset of nodes, representing flow hubs, and spreads gradually to the periphery, where only weak flows are present.

In the classical description of advection in continuous media, evolution of the concentration $u$ of passive particles in a given flow field $\vec v(\vec r)$ is described by the equation
${\partial u}/{\partial t} + \textrm{div} (\vec v u ) = 0$.
If flows are conserved, condition $\textrm{div} (\vec v )= 0$ should additionally hold.
What would be the analog of this advection equation for stochastic transport of particles by conserved flows on networks?

Let us consider a network of size $N$.
The network topology is determined by the adjacency matrix $\mathbf A$ whose elements are $A_{ij}=1$, if there is a link from node $j$ to node $i$,
and $A_{ij}=0$ otherwise.
Passive particles occupy network nodes and are transported with certain probabilities together with flows over the links that connect them.
Their stochastic advection corresponds to a Markov process (a random walk) and the evolution of the concentrations $u_{i}$ of the particles in network nodes is described by equation
\begin{align}
\frac{\partial u_{i}}{\partial t} = \sum_{j=1}^{N} \left ( \nu J_{ij} A_{ij} u_{j} - \nu J_{ji} A_{ji} u_{i}\right ).
\label{eq01}
\end{align}
It is important that, in the advection problem, the probability rate $\nu_{ij}$ for the transition from node $j$ to node $i$ is proportional to the intenstity $J_{ij}$ of the flow along the respective link, $\nu_{ij} = \nu J_{ij}$ and, furthermore, the flows are conserved.
Therefore, the total incoming flow in each node is equal to the total outgoing flow.
Thus, the condition
\begin{align}
\sum_{j=1}^{N} J_{ij}A_{ij} = \sum_{j=1}^{N} J_{ji}A_{ji},
\label{eq02}
\end{align}
should hold for any node $i$.

Note that, because of the conditions~(\ref{eq02}), the incoming and outgoing flows become correlated and hence the flows $J_{ij}$ cannot be arbitrarily assigned.
To construct the flow pattern on a network, one can proceed in the following way.
Suppose that $X_{i}$ is the total incoming flow in the node $i$, i.e.~$X_{i} = \sum_{j=1}^{N}J_{ij}A_{ij}$.
If all network links are identical in terms of their transportation capacities, it is natural to assume that the incoming flow $X_{i}$ is equally divided among all outgoing links of the node $i$.
Then, for a link from node $i$ to node $j$, we have 
$J_{ji} = {X_{i}}/{k_{i}^{\textrm{out}}}$
where $k_{i}^{\textrm{out}} = \sum_{l=1}^{N}A_{li}$ is the outgoing degree of node $i$
\footnote{
We consider only the networks where each node has at least one outgoing connection, so that $k_{i}^{\textrm{out}}\neq 0$.
}.

Generally, links may have different transportation capacities $w_{ij}$.
In this case, the flow is divided among the outgoing links according to their relative transportation capacities, so that
\begin{align}
J_{ji} = \frac{w_{ji}X_{i}}{\sum_{l=1}^{N}w_{li}A_{li}}.
\label{eq03}
\end{align}
In absence of external sources, flows $X_{i}$ passing through the node can be therefore found as solutions of equations
\begin{align}
\sum_{j=1}^{N} \left ( \frac{w_{ij}A_{ij}}{\sum_{l=1}^{N} w_{lj}A_{lj}} - \delta_{ij} \right ) X_{j} = 0.
\label{eq04}
\end{align}
Once they are known, flows $J_{ij}$ along the links can be obtained using Eq.~(\ref{eq03}).
It should be stressed that the flow pattern is a global property of a network and the flow $X_{i}$ in a given node may strongly change when perturbations in the network structure far from this node have occurred.

\begin{figure}[t]
\begin{center}
\includegraphics[width=0.9\hsize]{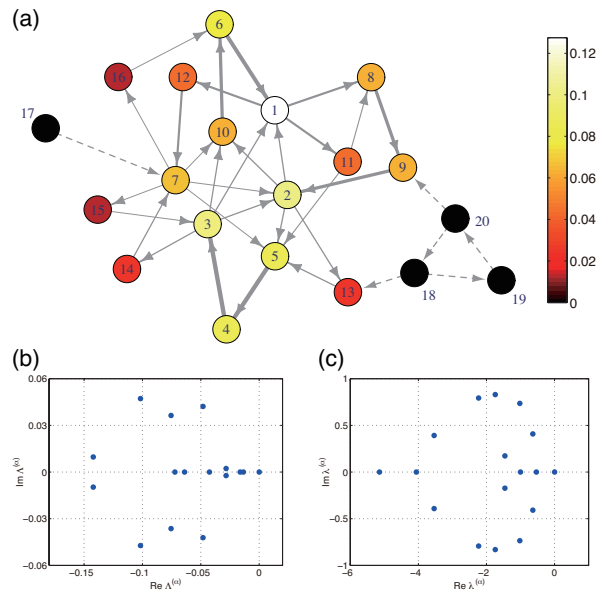}
\caption{
(a) Flow pattern in a network of size $N=20$.
Arrows represent directed connections.
The thickness of each arrow characterizes the flow $J_{ij}$ passing through the respective link.
Dashed arrows correspond to the links without flows.
The total flow $X_{i}$ is displayed by using the color code in the bar.
In panels (b) and (c), eigenvalue spectra $\Lambda^{(\alpha)}$ and $\lambda^{(\alpha)}$ of the advection matrix $\mathbf M$ and the Laplacian matrix $\mathbf L$ for the same network are shown.
}
\label{fig01}
\end{center}
\end{figure}

Unless otherwise specified, only networks with equal transportation capacities of the links will be considered below, so that $w_{ij} = 1$ in Eq.~(\ref{eq04}).
Moreover, we use the normalization $\sum_{i=1}^{N} {X_{i}}=1$.
It is convenient to enumerate nodes according to the flows $X_{i}$ which pass through them,
so that
$X_{1} \geq X_{2} \geq \cdots \geq X_{N}$
and the nodes with the smallest indices represent {\it flow hubs}.
Figure~\ref{fig01} shows an example of a network with its flow pattern.
As seen from this figure, flow hubs do not generally correspond to network hubs, i.e.~the nodes with the largest incoming or outgoing degrees
(compare nodes 1, 5 and 7).
Furthermore, there are nodes (17 to 20 in Fig.~\ref{fig01}(a)) where flows are absent.

An important property of considered advection processes is that they lead to equilibration of particles' concentrations in all network nodes (except for a subset of nodes through which flows do not pass).
Indeed, it can be easily checked that the uniform distribution $u_{i}=\textrm{const}$ is always a stationary state of Eq.~(\ref{eq01}) if flow conservation conditions~(\ref{eq02}) are satisfied.

By introducing an advection matrix $\mathbf M$ with elements
$M_{ij} = J_{ij} A_{ij} - \sum_{l=1}^{N}J_{li} A_{li} \delta_{ij}$,
and vectors $\vec u$ with component $u_{i}$,
Eq.~(\ref{eq01}) can be rewritten as
$\dot {\vec u} = \mathbf M \vec u$,
where, for simplicity, we choose $\nu = 1$.
Their general solution is given by
\begin{align}
\vec u(t) = \sum_{\alpha=1}^{N} c^{(\alpha)} \exp \left [ \Lambda^{(\alpha)}t \right ] \vec \phi^{(\alpha)},
\label{eq05}
\end{align}
where
$\Lambda^{(\alpha)}$ and $\vec \phi^{(\alpha)}$, are the eigenvalues and the eigenvectors of the advection matrix,
$\mathbf M \vec \phi^{(\alpha)} = \Lambda^{(\alpha)}\vec \phi^{(\alpha)}$ and the coefficients $c^{(\alpha)}$ are determined by initial conditions.

The spectrum of the advection matrix plays an important role in the evolution of a concentration pattern.
It can be straightforwardly checked that the advection matrix is negative semidefinite and, therefore,
real parts of all its eigenvalues are nonpositive, $\textrm{Re}\Lambda^{(\alpha)}\leq 0$.
The eigenvector with the zero eigenvalue corresponds to the stationary state which, as we have noted above, represents a uniform distribution.
Hence, Eq.~(\ref{eq05}) describes a relaxation process.
Note that the index $\alpha$ can always be assigned in such a way that
$\textrm{Re}\Lambda^{(1)}\leq \textrm{Re}\Lambda^{(2)}\leq \cdots \leq \textrm{Re}\Lambda^{(N)}$
and we have
$\Lambda^{(N)}=0$.

In addition to the advection matrix, it is also possible to define the Laplacian matrix of the same network with the elements $L_{ij} = A_{ij} - \delta_{ij} \sum_{i=1}^{N}A_{ij}$.
These two matrices - and, therefore, also there eigenvectors and eigenvalues - are generally different.
As an example, Figs.~\ref{fig01}(b) and (c) show spectra of the advection and the Laplacian matrices of the network in Fig.~\ref{fig01}(a).

\begin{figure}[t]
\begin{center}
\includegraphics[width=0.95\hsize]{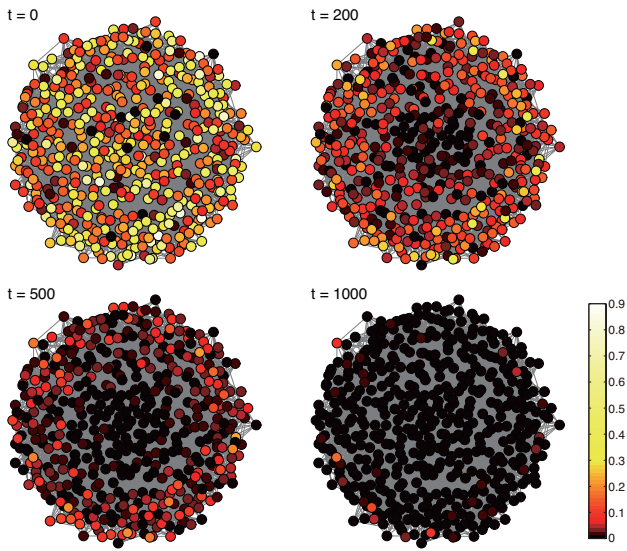}
\caption{
Mixing process in a scale-free flow network of size $N=500$ and mean degree $\langle k \rangle = 20$.
The color code shows deviations of the concentration from the uniform steady state.
}
\label{fig02}
\end{center}
\end{figure}

Figure~\ref{fig02} shows an example of the mixing process in a flow network starting from a random initial condition.
This scale-free network was generated by the preferential attachment algorithm~\cite{Barabasi1999}.
The direction of each link was randomly chosen under a restriction that each node has at least one incoming and one outgoing links.
The network size is $N=500$ and the mean degree (number of links per node) is $\langle k \rangle = 20$.
The simulation started from a random concentration distribution ($t=0$ in Fig.~\ref{fig02}).
In the visualization employed in Fig.~\ref{fig02}, network nodes with large passing flows $X_{i}$ (flow hubs) are located at the center
and the nodes with weak passing flows are in the periphery of the graph.
The equilibration first takes place in the center, at flow hubs ($t=200$).
It gradually spreads over the network ($t=500$).
At the final stage, periphery nodes become equilibrated ($t=1000$).

Such mixing equilibration behavior is general, it could always be seen in the numerical simulations for various networks.
As we show below, it can be explained by the localization of eigenvectors of the advection matrix $\mathbf M$.

According to Eq.~(\ref{eq05}), the initial distribution at time $t=0$ can be decomposed into the sum of contributions $c^{(\alpha)}$ corresponding to different eigenmodes $\alpha$ of the advection matrix.
As time goes on, first the contributions with large relaxation rates $| \textrm{Re} \Lambda^{(\alpha)} |$ should disappear and,
generally, at time $t=T$ only the contributions corresponding to $| \textrm{Re} \Lambda^{(\alpha)} | \lesssim T^{-1}$ would remain.

\begin{figure}[t]
\begin{center}
\includegraphics[width=0.97\hsize]{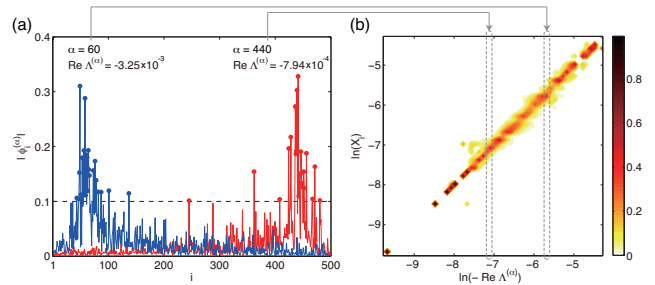}
\caption{
Localization of eigenvectors of the advection matrix for the scale-free network shown in Fig.~\ref{fig02}.
(a) Two eigenvectors for $\alpha = 60$ (blue) and $\alpha = 440$ (red).
Magnitudes $| \phi^{(\alpha)}_{i} |$ are displayed and large-deviation points where $| \phi^{(\alpha)}_{i} | \geq 0.1$ are marked by dots.
(b) Density plot of the large-deviation points (see the text).
}
\label{fig03}
\end{center}
\end{figure}

Eigenvectors $\vec \phi^{(\alpha)}$ of the advection matrix are localized on the network, as illustrated in Fig.~\ref{fig03}.
Figure~\ref{fig03}(a) displays two different eigenvectors with $\alpha = 60$ and $440$.
It can be seen that the eigenvector corresponding to the smaller $\alpha$ is localized on the subset of nodes with small indices $i$.
Because we enumerate the nodes in the order of the decreasing flows $X_{i}$, the nodes with the small indices are actually flow hubs.
On the other hand, at $\alpha=440$ the eigenvector is localized on a subset of nodes with high indices $i$ where flows $X_{i}$ are weak.

According to Fig.~\ref{fig03}(b), localization holds for all eigenmodes $\alpha$.
We have constructed this density plot in the following way:
For each eigenvector $\vec \phi^{(\alpha)}$, all nodes were divided into groups according to their flows $X_{i}$.
Each group contained the nodes with the flows $X_{i}$ within the window of width $0.1$ for the variable $\textrm{ln} (X_{i})$.
For each group, the numbers of the large-deviation nodes with $| \phi^{(\alpha)}_{i} | \geq 0.1$ was counted.
Furthermore, the variable $\textrm{ln} (| \textrm{Re}\Lambda^{(\alpha)} |)$ was divided into equal intervals of width $0.1$ and the number of the large-deviation nodes for all eigenvectors $\vec \phi^{(\alpha)}$ with $\alpha$ within the same interval were summed up.
The resulting relative numbers of the large-deviation node in each cell are displayed as a density plot in Fig.~\ref{fig03}(b).
One can see that large-deviation nodes are approximately located along the diagonal of the density map.
This means that, for each eigenmode $\alpha$, there is a characteristic flow $X_{\alpha}$ which specifies the large-deviation nodes
and $X_{\alpha}\simeq |\textrm{Re} \Lambda^{(\alpha)}|$.

\begin{figure}[t]
\begin{center}
\includegraphics[width=0.92\hsize]{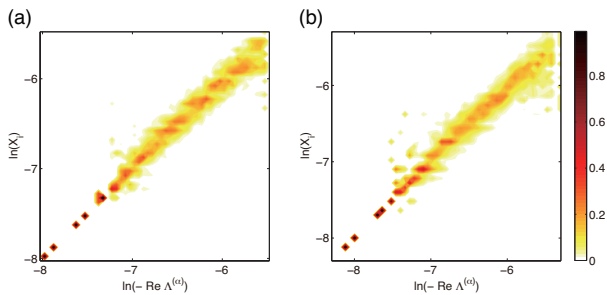}
\caption{
Density plots (see the text) for (a) the Erd\"{o}s-R\'{e}nyi random network of size $N=500$ and mean degree $\langle k \rangle = 20$
and (b) the same network with transportation capacities randomly drawn from the uniform distribution $0 < w_{ij} \leq 1$.
Counting intervals for these density plots were $\varDelta [ \textrm{ln} (-\textrm{Re} \Lambda^{(\alpha)}) ] = \varDelta [ \textrm{ln} (X_{i}) ] = 0.05$.
}
\label{fig04}
\end{center}
\end{figure}

Our investigations show that localization is not significantly sensitive to the topology and the size of random networks.
We could observe it for scale-free networks of different sizes $N$ and it was also present for Erd\"{o}s-R\'{e}nyi networks of size $N=500$, as illustrated by Fig.~\ref{fig04}(a).
Moreover, the condition $X_{\alpha}\simeq |\textrm{Re} \Lambda^{(\alpha)}|$ was always found to hold.

So far, only networks where all links have been identical in terms of their transportation capacities were considered.
However, our analysis can be straitforwardly extended to the situation when different transportation capacities $w_{ij}$ are assigned to the links.
In this case, flows $X_{i}$ can be computed from Eqs.~(\ref{eq04}) and flows $J_{ij}$ along the links are given by Eq.~(\ref{eq03}),
so that the respective advection matrix $\mathbf M$ is obtained.
As it turns out, the eigenvectors of such advection matrix are also localized on the subsets of nodes with some characteristic flows $X_{\alpha}$.
Figure~\ref{fig04}(b) shows the density plot for the advection matrix which corresponds to the same Erd\"{o}s-R\'{e}nyi network as in Fig.~\ref{fig04}(a), but with different transportation capacities randomly assigned to each of the links.
Thus, the localization effects are apparently universal and, therefore, the mixing process should have similar properties in various kinds of networks.

Our study reveals that there are essential differences between advection phenomena and diffusion of particles over networks.
Generally, a random walk is described by the master equation for a stochastic Markov process which has the same form as Eq.~(\ref{eq01}), but where the transition rates $\nu_{ij}$ may be arbitrarily chosen.
It leads to establishment of a steady state where final concentrations $u_{i}$ of particles in network nodes are different.
Only for diffusion, i.e.~if the transitions are symmetric and $\nu_{ij}=\nu_{ji}$, concentrations are equal in the steady state.
In the advection problem, even if the flows are allowed to pass only in one direction along a link, the steady state always represents a uniform distribution and this directly follows from the flow conservation condition~(\ref{eq02}).


Localization has previously been considered for network diffusion processes, where eigenvectors of the Laplacian matrix play an important role and the localization is determined by degrees of the nodes~\cite{McGraw2008, Nakao2010, Hata2013}.
In contrast to this, localization of eigenvectors of the advection matrix is determined by the flows passing through network nodes and, generally, flow hubs are different from network hubs.
Accordingly, equilibration of concentrations in advection phenomena starts in flow hubs and proceeds to the flow periphery of a network.

Moreover, there is also a difference with respect to the classical problem of advection by hydrodynamical flows,
described by equation ${\partial u}/{\partial t} + \textrm{div} (\vec v u ) = 0$.
If flows are turbulent, mixing takes place and a uniform state is eventually established~\cite{Shraiman2000}.
When hydrodynamical flows are stationary, there is no mixing and no relaxation to a uniform state.
Mixing, leading to equilibration of particle concentrations in our problem, is due to stochastic nature of transitions between the nodes.

Note that our study refers only to patterns of conserved circulating flows on the networks.
However, the analysis can be straightforwardly extended to the networks which include flow sources and sinks.
Our investigations were focused on the mathematical aspects and specific applications have not been considered here.
In the future, it may be interesting to apply the theory, for example, to the situations where infection or pollution spreading by conserved traffic flows is involved.

Financial support through the DFG  SFB 910 program in Germany,
through the Fellowship for Research Abroad, KAKENHI and the FIRST Aihara Project (JSPS),
and the CREST Kokubu Project (JST) in Japan is acknowledged.



\end{document}